\documentclass[12pt]{article}
\usepackage[english]{babel}
\usepackage[centertags]{amsmath}
\usepackage{amsfonts}
\usepackage{amssymb}
\usepackage{amsthm}
\usepackage{graphicx}
\usepackage{subfigure}
\usepackage{multicol}
\usepackage{multirow}

\theoremstyle{plain}
\newtheorem{thm}{Theorem}[section]

\theoremstyle{definition}

\theoremstyle{remark}

\numberwithin{equation}{section}

\numberwithin{equation}{section}


\begin{document}
\title{Modelling cross-border systemic risk in the European banking sector: a copula approach}
\author{Raffaella Calabrese$^1$ and Silvia Angela Osmetti$^2$\\\\
$^1$University of Essex\\
rcalab@essex.ac.uk\\ \\
$^2$ Universit\`{a} Cattolica del Sacro Cuore di Milano\\
silvia.osmetti@unicatt.it}
\date{}
\maketitle
\begin{abstract}
We propose a new methodology based on the Marshall-Olkin (MO) copula to model cross-border systemic risk. The proposed framework estimates the impact of the systematic and idiosyncratic components on systemic risk. Initially, we propose a maximum-likelihood method to estimate the parameter of the MO copula. In order to use the data on non-distressed banks for these estimates, we consider times to bank failures as censored samples. Hence, we propose an estimation procedure for the MO copula on censored data. The empirical evidence from European banks shows that
the proposed censored model avoid possible underestimation of the contagion risk.
\end{abstract}
\section{Introduction}

The 2007-2008 financial crisis has shown how a shock that originates in one country or asset class can quickly propagate to other markets and across borders. More importantly, it disclosed the major role of interconnectedness among banks in the propagation of financial distress. Interconnections, due to bilateral contractual obligations, exposure to
common risk factors and sudden collapses in market confidence, have grown dramatically in the
run-up to the crisis. While higher interconnectedness is a crucial means of efficient risk transfer, it
may also lead to contagious default cascades: an initial shock may propagate throughout the entire
banking system via chains of defaults follow highly dynamic patterns.

Direct and indirect linkages among banks arose as a key component of financial contagion in the European Union, as revealed first by the default of Lehman Brothers in September 2008, and then by the euro area sovereign debt crisis. In the Euro area, the cross-border exposures arose as a prominent issue with the European sovereign debt crisis in 2011 and 2012, where large exposure of many EU banks to stressed sovereigns were revealed by the European Banking Authority \cite{EBA}. In a broader perspective, correlated exposures have recently been shown to be a major source of systemic risk.

Given the importance of this research field, this paper is focused on systemic risk in the European banking sector.
By definition, systemic risk involves the financial system, a collection of interconnected
institutions that have mutually beneficial business relationships through which insolvency can quickly propagate during periods of financial distress \cite{Billio}. Systemic risk is mainly due to \emph{idiosyncratic} and \emph{systematic shocks} (\cite{De Bandt}, \cite{ECB}). The first ones are those which affect only the health of a single financial institution, while the latter
affect the whole economy, e.g. all financial institutions together at the same time. The component of systemic risk due to idiosyncratic shocks is also known as \emph{contagion risk} in the literature \cite{De Bandt}.

The main aim of this paper is to propose a new methodological approach for the analysis of systemic risk to model jointly idiosyncratic and systematic shocks. We propose to apply the copula approach to measure systemic risk between the banking sectors of two countries. To our knowledge, the only papers that previously applied copulae to assess banking system stability are \cite{Strae} and \cite{Weiss}. In other words, the approach is pretty novel to the area of banking and systemic risk.

The contributions of this paper are threefold. The first proposal of this paper is to apply the Marshall and Olkin (MO) copula for modelling systemic risk between two European countries. Since this is an extreme value copula, it is suitable to study the dependence between extreme events as bank failures. Moreover, since the MO copula shows an upper tail dependence, in order to apply it to systemic risk, we suggest to consider the distribution function (df) of time to default for each country as the marginal df of the MO copula. Hence, the dependence is stronger for high values of distress probabilities of banks. Finally, the most important advantage of the MO copula is that it has both an absolute continuous part and a singular part. Thanks to the singular component, we can assign a non-null probability to the event that two banks in two countries show the same probability of distress at the same time. In this way, the parameter estimate of the MO copula represents the impact of the aggregate shocks on the systemic risk. To our knowledge, this is the first method suggested in the literature that allows to estimate the contribution of the systematic component to the systemic risk.

The second contribution of this paper is to consider a maximum likelihood method in order to estimate the dependence parameter of the MO copula. This procedure overcomes the complexity given by the presence of both a continuous part and a singular part of the MO copula. In this work, we apply the suggested methodology to balance sheet data of European banking systems. Since some EU banking systems, e.g. Italy and Germany, are characterised by a large number of small banks, a measure of bank distress can be estimated for all banks using the balance sheet approach.
We pair up banks in two European countries in terms of their probabilities of bank distress that we estimate using the BGEVA model (\cite{Ca} and \cite{Mar}). In order to estimate the marginal cds of the MO copula, we use the empirical cumulative distribution function of time to default for each country.

The third innovative aspect of this work is to consider a censored sampling, i.e. the time to default for non-distressed banks is right censored. In this way all the information of non-distressed banks can be used to estimate the parameter of the dependence structure. Finally, we suggest a maximum likelihood method to estimate the parameters of the MO copula for censored sampling. To our knowledge, this is the first paper that applies the MO copula and the censored sampling to model systemic risk.

We apply the proposals of this paper to data on Italian, German and UK banks over the period 1995-2012. The European sovereign debt crisis of 2009 is included in the empirical analysis. At first, we estimate the probability of distress for banks in each country
using the BGEVA model (\cite{Ca} and \cite{Mar}) on a set of bank specific factors addressed by the CAMELS framework (e.g. \cite{Arena}). In order to represent the economic cycle, we include also some macroeconomic variables in the BGEVA model. The estimates so obtained are used to pair up banks in two countries. In the country with the higher number of banks, we consider only the banks with a higher bank distress probability.

We compare the MO copula with the copula models used in the literature \cite{Weiss}, such as the Gaussian copula, the Gumbel copula and a mixture of the Frank, Clayton and Gumbel copula. An important result of this empirical analysis is that the estimate of the upper tail dependence in the MO copula is higher thanks to the singular component. Moreover, according to different goodness-of-fit measures, the MO copula is the model that best fits the data. Finally, when we apply censored techniques to the data, we obtain that the impact of the systematic component on the systemic risk increases.

We organise the paper as follows. The next section describes the literature review. Section 3 explains our methodological proposal. Section 4 describes the dataset and reports the main results on cross-border systemic risk. Finally, the last section
contains some concluding remarks. In the appendix, we report the score functions to obtain the maximum likelihood estimator of the parameter of the MO copula for censored sampling.

\section{Literature review}
\cite{ECB} and \cite{De Bandt} have identified two 'forms' of systemic risk, namely contagion risk, the risk that widespread imbalances that have built up over time unravel abruptly, and the risk of macro shocks causing simultaneous failures. It has been observed that many banking crises have occurred in conjunction with cyclical downturns or other aggregate shocks, such as interest rate increases, stock market crashes or exchange rate devaluations (see e.g. \cite{Acharya}).

The empirical literature on financial contagion has largely been divided along two different strands: one area of research has focused on capturing contagion using financial market data, see e.g. \cite{Cappiello}, \cite{Mang},\cite{Gropp}, \cite{Hartmanna},\cite{Hartmannb}, \cite{Longin}, \cite{Polson} and \cite{White}. A second strand has focused on banks' balance sheet data with the aim of analysing the potential effects on the network of the financial institutions if one or more of them are assumed to ecounter problems captured by balance sheet data, see e.g. \cite{Boss}, \cite{Degryse}, \cite{furfine}, \cite{Mistrulli}, \cite{Upper}, \cite{van Lelyveld} end \cite{Sormaki}.

Different methodologies have been applied to analyse contagion effects.  In some studies, contagion is presumed to be present if negative abnormal returns can be detected in the post-crisis period after the event that is supposed to be causing the bank panic (see e.g. \cite{Akhigbe}, \cite{Groppb}, \cite{Kabir}). Some authors have tried to use extreme value theory to estimate the number of joint occurrences of extreme events in the left tail of a bivariate series in order to isolate contagion effects across banks (\cite{Groppb} and \cite{Gropp}). Finally, the changes of dependencies between banks can be directly assessed by a copula-based approach (\cite{Vries}; \cite{Strae} and \cite{Weiss}).

\cite{Weiss} captures the changes in the dependence structure of abnormal bank returns by analysing the changes in the parametric form and the parameters of various copulae. In particular, he analyses changes in the dependence structure of banks around bailout announcements. To cover a maximal variety of tail dependence structures, \cite{Weiss} consider a convex combination over time of the Student's t, Frank, Clayton and Gumbel copula. In order to decide which convex mixture of parametric copulae is best suited for modelling the dependence structure, the author first estimated each possible mixture of three or four parametric copulae and computed the corresponding Akaike's Information Criterion. For the logarithmic stock returns of German banks, the Clayton-Frank Gumbel mixture is the best choice according to Akaike's criterion. Finally, \cite{Vries} suggests the Gumbel copula with Pareto marginal dfs as a joint distribution of the returns on syndicated loans in order to obtain heavy tailed marginal dfs, positive correlation and asymptotic independence.

We highlight that all the previous copulae are continuous, this means that the impact of the aggregate shocks on systemic risk could be underestimated. We overcome this drawback by applying the MO copula. Furthermore, another disadvantage of the copula-based approaches analysed above is that they use financial market data. Since we would apply the MO copula to the European banks and since most of them are small banks, we do not have market data for them. For this reason, we use banks' balance sheet data.

\section{A new method for modelling systemic risk}

In this work we propose to model the dependence structure of cross-border bank failures by the copula approach.
The concept of copula represents a flexible method since it does not require parametric assumptions on the marginal components (\cite{Ne}, and \cite{Fi}). In this way, a general class of distributions can be expressed through a simple model specification.

There are several advantages in applying the copula approach to systemic risk. The first one is that copula function is a suitable model to represent the dependence between rare events. Since the sample percentage of bank failure is much lower than 5\%, it could be classified as rare event. Furthermore, the copula model accounts for non-linear dependence and upper tail dependence. It is often shown in the literature, e.g. \cite{Weiss}, that contagion phenomena
cannot be captured by simple linear approaches like, e.g., regression analysis. Hence, capturing the tail dependence feature is essential for accurately assessing systemic risk, as shown in \cite{Vries} and in \cite{Strae} for bank stock returns. Finally, since we do not need to specify the marginal distributions, only the characteristics of the dependence structure are important.

As there are many copula families available \cite{Ne}, the appropriate copula for systemic risk is the one which best captures dependence features of bank failures. In order to represent the characteristics of bank distress
above analysed, we suggest to apply an extreme value copula with tail dependence, as explained in the following section.

\subsection{Copulae and tail dependence}

Every bivariate and multivariate cumulative distribution function (cdf)
$F$ and therefore every survival function $\overline{F}$ can be
treated as the result of two components: the marginal
distributions and the dependence structure. The copula describes
the way that the two marginal distributions are put together into the bivariate cdf or survival function.

In mathematical terms, a bivariate copula is a function $C:I^2\rightarrow I$, with
$I^2=[0,1]\times[0,1]$ and $I=[0,1]$, that satisfies
all the properties of a cdf. In particular, it
is the bivariate cdf of
a random variable (rv) $(U,V)$ with uniform marginal rvs
in [0,1]
\[C(u,v)=P(U\leq u,V\leq v),\;\;\;\;0\leq u\leq 1\;\;\;\;0\leq v \leq 1.\]

To better understand the copula model we consider the Sklar's
theorem \cite{Sk}.
\begin{thm}[Sklar]\label{theorem}
Let $(X,Y)$ a bivariate random variable with joint cumulative distribution
function $F_{X,Y}(x,y)$ and marginals $F_{X}(x)$ and $F_{Y}(y)$. It exists
a copula function $C:I^2\rightarrow I$ such that
$\forall x,y\in \mathcal{R}$
\begin{equation}\label{sklar}
F_{X,Y}(x,y)=C(F_{X}(x),F_{Y}(y))
\end{equation}
If $F_{X}(x)$ and $F_{Y}(y)$ are continuous functions then the copula $C(\cdot)$
is unique. Conversely, if  $C(\cdot)$ is a copula function and $F_{X}(x)$ and
$F_{Y}(y)$ are marginal cdfs, then the
$F_{X,Y}(x,y)$ in $(\ref{sklar})$ is a bivariate cdf.
\end{thm}

If $F_X(x)$ and $F_Y(y)$ are continuous cdfs, from $(\ref{sklar})$ the copula function results

\begin{equation}\label{defcopula}
C(u,v)=F_{X,Y}(F_X^{-1}(u),F_Y^{-1}(v))
\end{equation}
where $u=F_{X}(x)$ and $v=F_{Y}(y)$ are the cdfs $F_X(\cdot)$ and $F_Y(\cdot)$, respectively.

Analogously, if $\overline{F}_{X}(x)$ and $\overline{F}_{Y}(y)$ are continuous survival distribution functions (sdf) then the survival copula function (scf) $\widehat{C}:I^2\rightarrow I$ is
\begin{equation}\label{aggiunta}
\widehat{C}(\overline{F}_{X}(x),\overline{F}_{Y}(y))=P(X>x,Y>y)=\overline{F}_{X,Y}(x,y),\end{equation}
where $\overline{F}_{X,Y}(x,y)$ is the bivariate sdf.

Thus, a copula captures the dependence structure between the marginals.

A pivotal characteristic for analysis systemic risk is the upper tail dependence \cite{Trivedi}. An upper tail dependence parameter is defined as
$\chi_u$ is
\begin{equation}\label{taildep}\chi_u=\lim_{u\rightarrow 1^{-1}}P[X>F_X^{-1}(u)|Y>F_Y^{-1}(u))]=\lim_{u\rightarrow 1^{-1}}P[Y>F_Y^{-1}(u)|X>F_X^{-1}(u))].\end{equation}

Higher is the value of $\chi_u\in(0,1]$, higher is the level of upper tail dependence. Analogously, the lower tail dependence parameter $\chi_l$ can be defined. A given copula family is characterised by a given values of lower and upper tail dependence parameters \cite{Ne}.

\subsection{The Marshall-Olkin copula}

We suggest to use the Marshall and Olkin copula (MO copula) to model the dependence structure between the times to default of banks in two countries. The bivariate Marshall and Olkin distribution is used in reliability analysis to model jointly failure times of two components in a system when the failure is due to both idiosyncratic shocks, given by the characteristics of the components, and shocks common to both the components. The Marshall and Olkin copula models the dependence structure of the namesake probability distribution.

The main advantage of our suggestion is that the dependence structure of times to bank distress could be due to both idiosyncratic and systematic shocks. As explained in Section 2, the literature shows that both the components are important to model systematic risk.

Cuadras and Aug\'{e} propose the MO copula function in 1981 (see \cite{Ne}).
In the case of two exchangeable marginal rvs $X$ and $Y$, the MO copula is defined as
\begin{equation}
\label{MO copulaOPULA}
C(u,v)=uv\min(u^{-\theta},v^{-\theta})
\end{equation}
where $\theta \in [0,1]$ represents the intensity of the (positive) relationship between the marginal cdfs. If $\theta=0$ then the rvs $X$ and $Y$ are stochastically independent and the MO copula becomes $C(u,v)=uv$.
If $\theta=1$ then there is a perfect positive association between the rvs $X$ and $Y$ and the MO copula becomes $C(u,v)=min(u,v)$. Furthermore, the MO copula copula is an extreme
value copula with a upper right tail dependence where $\theta$ is the upper tail dependence parameter $\chi_u$.

An important characteristic of the MO copula (\ref{MO copulaOPULA}) is that it has an absolute continuous part and a singular part (\cite{Ne} and \cite{Osmetti}). Thanks to the singular part, we can assign a non-null probability to the event that the probabilities of failure of two banks located in two countries are equal at the same time. Hence, the MO copula can be considered as a linear combination of the absolute continuous part $C_a$ and the singular part
$C_s $
\begin{equation}\label{cdfcopula}
C(u,v)=\frac{2-2\theta }{2-\theta }C_a (u,v)+\frac{\theta
}{2-\theta }C_s (u,v)
\end{equation}
where $C_s (u,v)=[\min (u^\theta ,v^\theta )]^{\frac{2-\theta
}{\theta }}$ for $u=v$ and $C_a(u,v)$ for $u\neq v$ is:
\[C_a(u,v)=\frac{2-\theta}{2-2\theta}[uv\min(u^{-\theta},v^{-\theta})]-\frac{\theta}{2-2\theta}C_s(u,v).\]
As explained in Section 2, the systemic risk is due to both the idiosyncratic and the systematic shocks. The first ones are mainly characterised by banks' characteristics, the latters represent characteristics common to both the countries, such as macroeconomic conditions. If the upper tail dependence parameter $\theta$ of the MO copula is higher than 0.5 then systematic shocks are more important than idiosyncratic shocks to explain systemic risk. On the contrary, if $\theta<0.5$ the impact of idiosyncratic shocks on systemic risk is higher than that of systematic shocks.

\subsection{A new parameter estimation method}

The widely used method to estimate the distribution function in $(\ref{sklar})$ is the maximum likelihood
method \cite{Jo}. If the number of the cdf parameters is high or no information about the functional form of the marginal cdfs, the Canonical Maximum Likelihood (CML)(\cite{Jo} and \cite {Xu}) method is applied.
In particular, the CLM is a two-step semiparametric estimation approach: in the first step the marginal cdfs are estimated by the empirical cdf, in the second step the copula parameters are estimated by the maximum likelihood method. \cite{Ko} and \cite{Du} obtain that the CML shows the best performance on both simulated and empirical financial data.

In this section we suggest to apply the CLM to estimate the MO copula. In order to apply the CLM, we need to compute the probability density function of the MO copula. At first, we define the measure
$\mu $ for each $B\in B_2^+ $
\begin{equation} \label{measure}
\mu(B)=\mu _2(B)+\mu _1 \left(B\cap \left\{ x:(x,x)\in R _2^+
\right\} \right)
\end{equation}
for each $B \in B_2^+$ where $\mu _2 $ is a 2-dimensional Lebesgue measure, $B_2^+ $ is the Borel $\sigma
-$algebra in $R _2^+ $ and $\mu_1 $ is the Lebesgue measure on
the real line. Analogously to \cite{PrSu}, we obtain that the MO copula (\ref{MO copulaOPULA}) is absolutely continuous with respect to the
measure $\mu$.

We compute the derivative of the cdf $(\ref{cdfcopula})$ and we obtain the probability density function of the MO copula with respect to the measure (\ref{measure})
\begin{equation} \label{dencop}
c_\theta (u,v)
 =\left\{\begin{array}{lll}

 (1-\theta )\displaystyle\frac{1}{uv}C_\theta (u,v)& \{u>v\} \cup \{u<v\} \\
\\

 \theta \displaystyle \frac{1}{u}\mbox{ }C_\theta(u,v)& u=v \\

\end{array}  \right.
\end{equation}
with $0\le v\le 1$, $0\le u\le 1$ and $0<\theta <1$.

We can now apply the CLM. In the first step, we consider the empirical cdf as a non-parametric estimator of the cdf of the time to bank failure for each country $\hat{u}_i =\hat
{{F}}_X (x_i)$ and $\hat {v}_i
=\hat
{{F}}_Y (y_i)$. In the second step, we estimate the parameter $\theta\in (0,1)$ of the MO copula by maximizing the
conditional likelihood function
\[\hat {\theta}=arg \max L(\theta \vert \underline{\hat {u}},\underline{\hat {v}})\]
where
\begin{equation}\label{con}
L(\theta \vert \underline{\hat {u}},\underline{\hat {v}})=\prod_{i=1}^n c_{\theta}(\widehat{u}_i,\widehat{v}_i)\propto
(1-\theta )^{n_1 +n_2 }\theta ^{n_3 }\prod\limits_{i=1}^n
{C_\theta (\hat {u}_i ,\hat {v}_i )}.
\end{equation}
$n_1$, $n_2$ and $n_3$ are the number of observations such that
$n_1=\sharp\{\hat {u}_i <\hat {v}_i\}$, $n_2=\sharp\{\hat {u}_i >\hat {v}_i\}$ and $n_3=\sharp\{\hat {u}_i =\hat {v}_i\}$.
Hence, the maximum likelihood
estimator of $\theta$ is
\[\hat {\theta }=(1+\exp (-\hat {\psi }))^{-1}\] with
\begin{equation}\label{solcomp}\nonumber
\hat {\psi }=-\ln \left[ {\frac{n-2n_3 -S_{\min } +\sqrt
{n^2+S_{\min }^2 -S_{\min } (2n-4n_3 )} }{2n_3 }} \right]
\end{equation}
with $n_3>0$ and $S_{\min } =\sum\limits_{i=1}^n {\min } (-\ln
(\hat {u}_i ),-\ln (\hat {v}_i ))$ (see \cite{Osmetti} for details).
\subsection{Censored time of bank failure}

The method suggested in the former section to estimate the MO copula allows to use only the information of banks in distress that represents a very low percentage of the sample. In order to use the important information of most of the banks that are not in distress, we suggest to apply a type I censored sampling on the right that consists in stopping the observation of banks conditions at time $t^*$ (the highest observed time to bank default).

\begin{figure}\begin{center}\label{figurea}
\includegraphics[width=0.7\columnwidth]{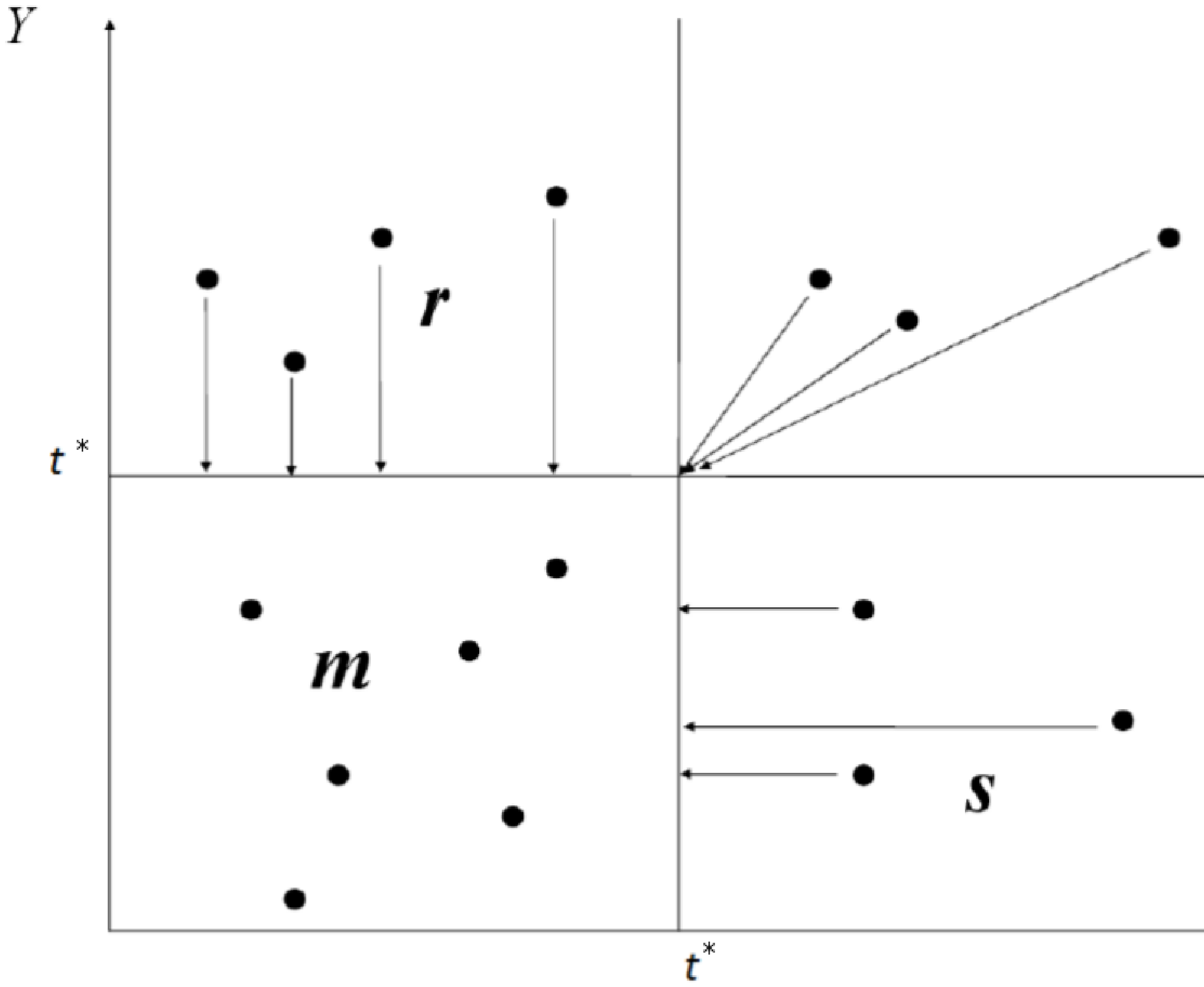}

\caption{Sample censored data}\end{center}
\end{figure}

To pair up banks located in two countries, we suggest to order banks in each country on the basis of their risk of distress. We define $m=\sharp\{x_i\leq t^* \cap y_i\leq t^*\}$ the number of bank pairs with both the banks from the two countries in distress.
Furthermore, we define $r=\sharp\{x_i \le t^* \cap y_i> t^*\}$ the number of banks in distress in the first country and not is distress in the second country and $s=\sharp\{x_i>  t^* \cap y_i\le t^*\}$ the number of banks not in distress in the first country and in distress in the second country.
This means that $n-m=\sharp\{x_i> t^* \cap y_i> t^*\}+r+s$ is the number of pairs where at least one bank of the two countries is not in distress. In order to apply a type I censored sampling, we assign $t^*$ to the time to default for non-distressed banks, as shown in the Figure $\ref{figurea}$.

In order to apply the type I censored sampling, we need to modify the CLM procedure suggested in the previous section. In the first step we estimate the marginal cds using the Kaplan-Maier estimator\footnote{The Kaplan-Maier estimator is used to estimate the cdf for a censored sample (see \cite{Kaplan}).}: $\hat {u}_i =\hat
{ {F}}_X (x_i)$, $\hat {v}_i =\hat
{ {F}}_Y (y_i)$.
Then, in the second step we maximize the conditional likelihood function of the copula.
Let $(\Delta^X,\Delta^Y)=(I_{\{X\le t^*\}})(x)$, $I_{\{Y\le t^*\}}(y))$, $\overline{\Delta}^X=1-\Delta^X$ and
$\overline{\Delta}^Y=1-\Delta^Y$, where $I_A(\cdot)$ is the
indicator function of the set $A$. Following \cite{Ow}, we compute
the conditional likelihood function for the copula
\begin{eqnarray}\nonumber
l(\theta|\hat{F} _X ,\hat{F}_Y )&=&\sum_{i=1}^n{\ln
[c_{\theta}(\hat{F}_X (x_i),\hat{F}_Y (y_i ))} ]^{\Delta_i^X
\Delta _i^Y }+\sum_{i=1}^n {\ln [C^1_{\theta} (\hat{F}_X (x_i
),\hat{F}_Y (y_i ))} ]^{\overline{\Delta} _i^X \Delta_i^Y
}+\\ &+&\sum_{i=1}^n {\ln [C^2_{\theta} (\hat{F}_X (x_i
),\hat{F}_Y (y_i ))} ]^{\overline{\Delta} _i^Y
\Delta_i^X}+\sum_{i=1}^n {\ln[C_{\theta}(\hat{F}_X (x_i
),\hat{F}_Y(y_i))} ]^{\overline{\Delta}_i^X
\overline{\Delta}_i^Y}\label{eq4}
 \end{eqnarray}
where $c_{\theta}(u,v)$ is the copula density defined in (\ref{dencop}),  $C^1_{\theta} (u,v)=\frac{\partial C_{\theta}(u,v)}{\partial v}$ and $C^2_{\theta}
(u,v)=\frac{\partial C_{\theta}(u,v)}{\partial u}$.\\

The maximum likelihood estimator of $\theta$ for type I censored data is

\begin{equation}\label{estimator}
\hat {\theta }_c =(1+\exp (-\hat {\psi }_c
))^{-1}
\end{equation}
with
\begin{equation}\label{phi}
\hat {\psi }_c =-\ln \left[ {\frac{m+r+s-2m_3 -S_{\min } +\sqrt
{(m+r+S_{\min } -2m_3 )^2+4m_3 (m+r+s-m_3 )} }{2m_3 }} \right]
\end{equation}
where $m_1=m-\sharp \{\hat {u}_i\geq \hat {v}_i\}$, $m_2=m-\sharp \{\hat {u}_i\leq \hat {v}_i\}$, $m_3=m-m_1-m_2$
and \[S_{\min } =\sum_{i=1}^m \min (-\ln (\hat {u}_i
),-\ln (\hat {v}_i ))+\sum_{i=1}^r[-\ln (\hat {u}_i )]+\sum_{i=1}^s [-\ln (\hat {v}_i
)]
+(n-m-r-s)t^*.\]

The maximum likelihood estimator (\ref{estimator}) is the unique and acceptable solution of this optimization problem (see Appendix \ref{App1} for details).

\section{Empirical results}
\subsection{Dataset}
The empirical analysis is based on annual data for the period 1995-2012 for Italian, German and UK banks. The data are from  Bankscope, a comprehensive database of balance sheet
and income statement data for individual banks across the world provided
by the private company Bureau Van Dijk. The time horizon and the geographic area are important for the European sovereign debt crisis of 2009. We choose to analyse the cross-border bank interdependence between Italy, German and UK since their banking systems are quite different. For example, most of the Italian and German banks are quite small and they are cooperative or savings banks (around 90\% in Germany). In UK the average bank size is larger and
there are not traditionally regional or state banks and only one cooperative bank.

All the three banking systems came under pressure during the financial and the sovereign debt crisis. UK banks were significant exposed to toxic assets which originated in US, Italian and German banks less. On the contrary, the impact of the sovereign debt crisis was stronger on the Italian and the German banking systems, even if the stability of the German system has been achieved in the short run in large part through substantial government support measures.

 In order to pair up banks located in two countries, in the previous section we proposed to order the banks in each country on the basis of their risk of failure. In particular, we apply the BGEVA model  (\cite{Ca},\cite{Mar}) to estimate the probability of default for each bank in a given country. This method is a semiparametric scoring model and it is particularly suitable for very small
 number of defaults in the sample. The early warning indicators of bank failures in the literature can be divided into two sets: those that are bank specific, i.e. the financial ratios associated with the CAMEL rating system (\cite{Arena}) and macroeconomic factors that affect all banks (\cite{Ca2}).

In order to measure the severity of multicollinearity we have computed the Variance Inflation
Factor (VIF) for each explanatory variable. Firstly, we consider 22
independent variables and we remove those with a VIF higher than 5,
so we obtain 18 covariates: Total Assets, Loan Loss Reserve over Gross Loans, Equity over Total Assets, Return on Average Assets (ROAA), Return on Average Equity (ROAE),
Net Loans over Total Assets, Liquid Assets over Cust\& ST Funding, Interbank Assets over Interbank Liabilities, Liquid Assets over Tot Dep \& Bor, Tier 1 Ratio, Total Capital Ratio, Equity over Liabilities, Equity over Net Loans, Net Interest Margin, Growth Rate of GDP, Inflation Rate, Unemployment Rate and Interest Rate.

All data are available for 1,802 German banks, 602 Italian banks and 265 UK banks. These sample sizes are coherent with the characteristics of the banking systems of these countries analysed above.

\subsection{Estimation results and goodness-of-fit measures }
As explained in the previous section, we apply the BGEVA model to banks of each country in order to estimate the default probabilities. These estimates are used first to order banks in each country and then to pair up banks located in two countries. To apply a bivariate copula, the numbers of banks in both the countries need to be the same. Hence, for the country with a higher number of banks, we consider only the banks with higher total assets, since systemic risk is more important for them.
Afterwards, we use the empirical cdfs of the time to default for each country as marginal cdfs of the MO copula. Finally, we apply both the methods suggested in Section 3.3. and 3.4 to estimate the parameter $\theta$ of the MO copula.

We compare the MO copula with the copula models used in the literature (\cite{Rod} and \cite{Weiss}), such as the Gaussian copula, the Gumbel copula and a finite mixture of the Frank $C_F$, Clayton $C_C$ and Gumbel $C_G$ copulae ($F+C+G$)

\[C(u,v)=\pi_F C_F(u,v;\alpha)+\pi_C C_C(u,v;\gamma)+(1-\pi_F-\pi_C) C_G(u,v;r)\]
with weights $0\leq\pi_i \leq 1$ for $i=F,C,G$. The MO, Gumbel and the mixture of copulae display asymptotic tail dependence and asymmetry, while the Gaussian copula is symmetric without tail dependence. The parameter $-1<\rho<1$ of the Gaussian copula represents the linear correlation coefficient. Furthermore, the parameter $r>1$ of the Gumbel copula is a measure of positive association and represents the intensity of the upper tail dependence $(\chi_u=2-2^{1/r})$. The Frank copula is a symmetric copula and it shows positive dependence for $\alpha \in (0,+\infty)$, negative
dependence for $\alpha \in (-\infty,0)$ and independence for $\alpha=0$. The tail dependence in this copula model is null. Finally, the Clayton copula shows also a positive dependence. Its parameter $\gamma$ represents the intensity of the lower tail dependence $(\chi_u=2^{-1/\gamma})$.

 Finally, the mixture of the Frank, Clayton and Gumbel copulae can display both the lower tail dependence, for the Clayton copula, and the upper tail dependence, for the Gumbel copula.

\begin{table}\caption{Copulae parameters estimates}\label{tab1}
\begin{center}\begin{tabular}{|c|c|c|c|}

\hline
\emph{Copula} & \emph{IT-UK} & \emph{IT-DE} & \emph{UK-DE} \\
\hline
Gaussian    & $\hat{\rho}=0.25$                     & $\hat{\rho}=0.30$                      &$\hat{\rho}=0.27$\\\hline
Gumbel      & $\hat{r}=1.30$                        & $\hat{r}=1.40$                         &$\hat{r}=1.37$ \\\hline
\multirow{4}{*}{$F+C+G$}  & $\hat{\pi}_F=0.31$, $\hat{\pi}_C=0.15$      & $\hat{\pi}_F=0.21$, $\hat{\pi}_C= 0.14$   & $\hat{\pi}_F=0.25$, $\hat{\pi}_C=0.13$\\
 & $\hat{\alpha}=0.01$    & $\hat{\alpha}=0.0004$     & $\hat{\alpha}=0.0003$ \\
 & $\hat{\gamma}=0.23$ &$\hat{\gamma}=0.26$ &$\hat{\gamma}=0.25$\\
 & $\hat{r}= 1.33$                       & $\hat{r}=1.45$                         & $\hat{r}=1.41$\\\hline
MO       & $\hat{\theta}=0.37$                   & $\hat{\theta}=0.55 $                   & $\hat{\theta}=0.45 $\\

\hline
\end{tabular}\end{center}
\end{table}

Table 1 shows the results obtained for different copula models. The linear correlation coefficient estimate $\rho$ of the Gaussian copula is close to zero for all the pairs of countries. This result could be due to the fact that the Gaussian copula displays only a linear
dependence and not a tail dependence. The latter is what we expect in the data.

To verify this expectation we apply a Gumbel copula that shows upper tail dependence and a mixture of copulae that displays both upper and lower tail dependence. Since the parameter $\hat{r}$ is higher than 1 for all the three pairs of countries, this means that there is upper tail dependence. The intensity of this dependence is quite low since all the values of $r$ are close to 1.

In agreement with the expectations, the Gumbel copula shows the highest weight in the mixture model for all the pairs of countries ($\pi_G$=0.62 for IT-UK, $\pi_G$=0.65 for IT-DE and $\pi_G$=0.54 for UK-DE). We use equation (\ref{taildep}) to compute the upper tail dependence parameter. We obtain $\chi_u$=0.316 for IT-UK, $\chi_u$=0.365 for UK-DE and $\chi_u$=0.387 for IT-DE. This means that the intensity of the upper tail dependence in the mixture model is still low. We highlight that the orderings of the upper tail dependence parameter estimates in both the mixture and the Gumbel copulae are the same. Furthermore, these orderings correspond to the one of the linear correlation coefficients in the Gaussian copula. From this ordering the systemic risk for IT-DE results higher than  that for DE-UK that is finally higher than the one for IT-UK. This result is in line with the expectations and with Gropp et al. (2009)'s outcomes. Gropp et al. (2009)  estimate the contagion directions of banks that experience a large shock on the same day. They obtain a strong bilateral relationship between Italy and Germany and a weak bilateral contagion between UK and Germany.

Finally We apply the MO copula. Its parameter $\theta$ represents the upper tail dependence parameter. From Table 1 we note that the MO model suggested in this paper is that it shows an higher tail dependence than the previous copula models. The tail association between the distressed banks in UK and Germany is medium-high ($\chi_u$=0.55), the one between the Uk and the German banking systems is medium-low ($\chi_u$=0.45).

The higher upper tail dependence of the MO copula could be due to include a singular part in the model in order to assign a non-null probability to the event that banks in two countries are in distress at the same time with the same probability. In this way, we can accurately estimate the systematic component of systemic risk. On the contrary, in the Gumbel and in the mixture model this component could be underestimated, as data show.

Given the idiosyncratic characteristics of banks, if $\theta$ is higher than 0.5 the systematic component is more important than the idiosyncratic one to explain systemic risk. Since Italian and German banks are under the same monetary policy of the European Central Bank, it is coherent that the most important component of systemic risk is the systematic one ($\theta>0.5$). This component becomes less important if the two banking systems are under two different monetary policies. Figure \ref{fig: copula countour} shows the estimated MO copula function and its contour levels for IT-UK, UK-DE and IT-DE.
\begin{center}\label{fig: copula countour}
\begin{figure}\caption{Mo copula and countour lines estimate for IT-UK ($\theta$=0.37)}
\includegraphics[width=0.83\columnwidth]{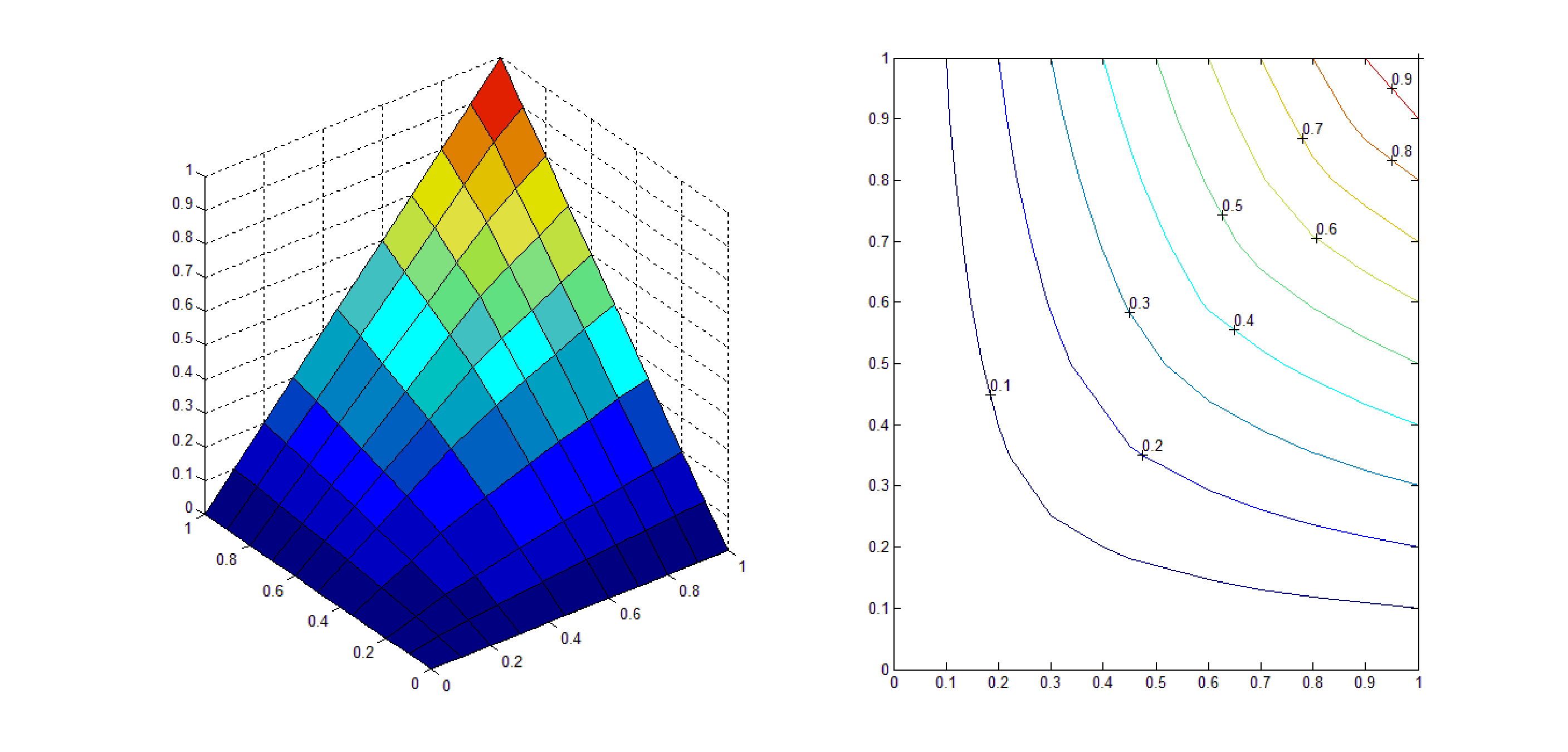}
\caption{Mo copula and countour lines estimate for UK-DE ($\theta$=0.45)}
\includegraphics[width=0.84\columnwidth]{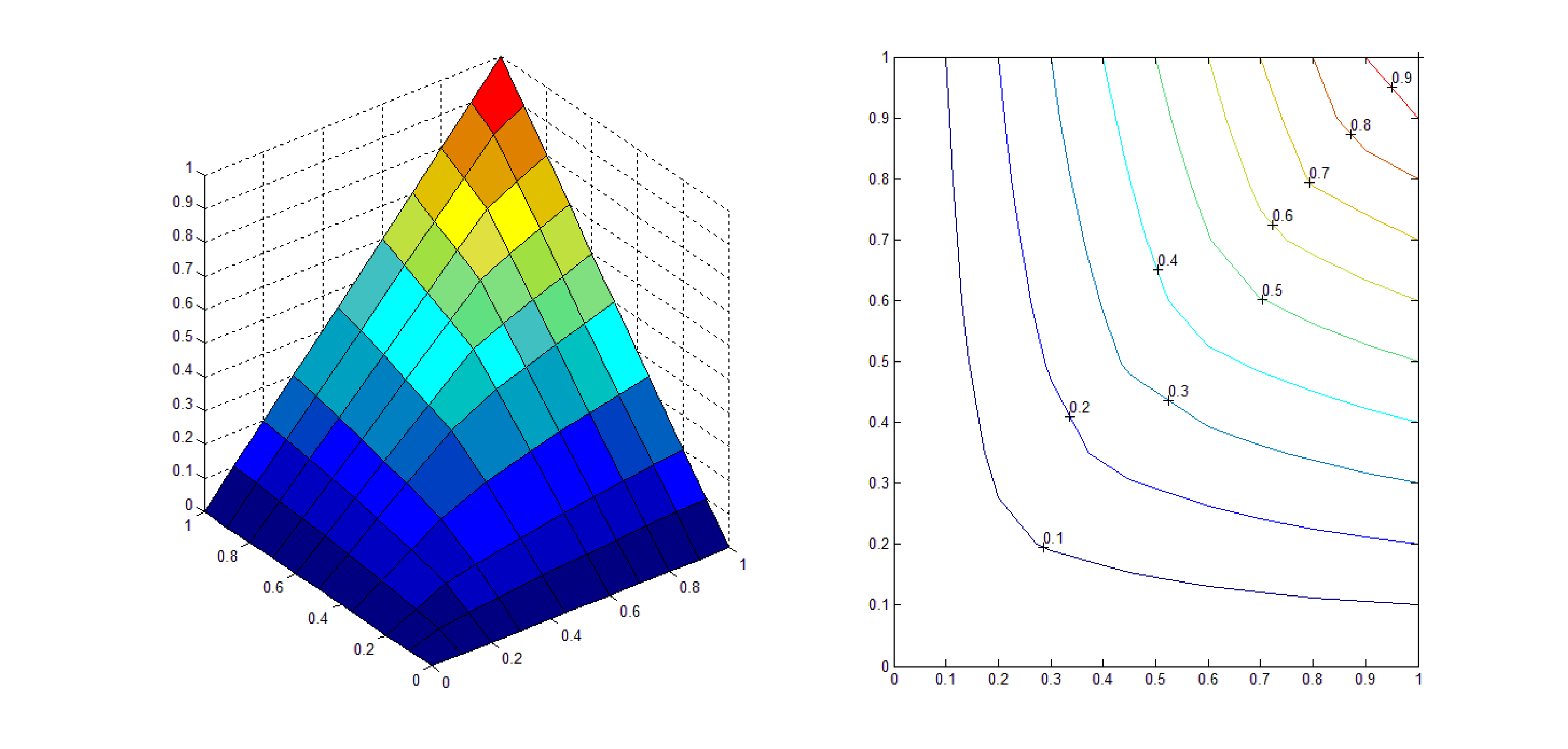}
\caption{Mo copula and countour lines estimate for IT-DE ($\theta$=0.55)}
\includegraphics[width=0.85\columnwidth]{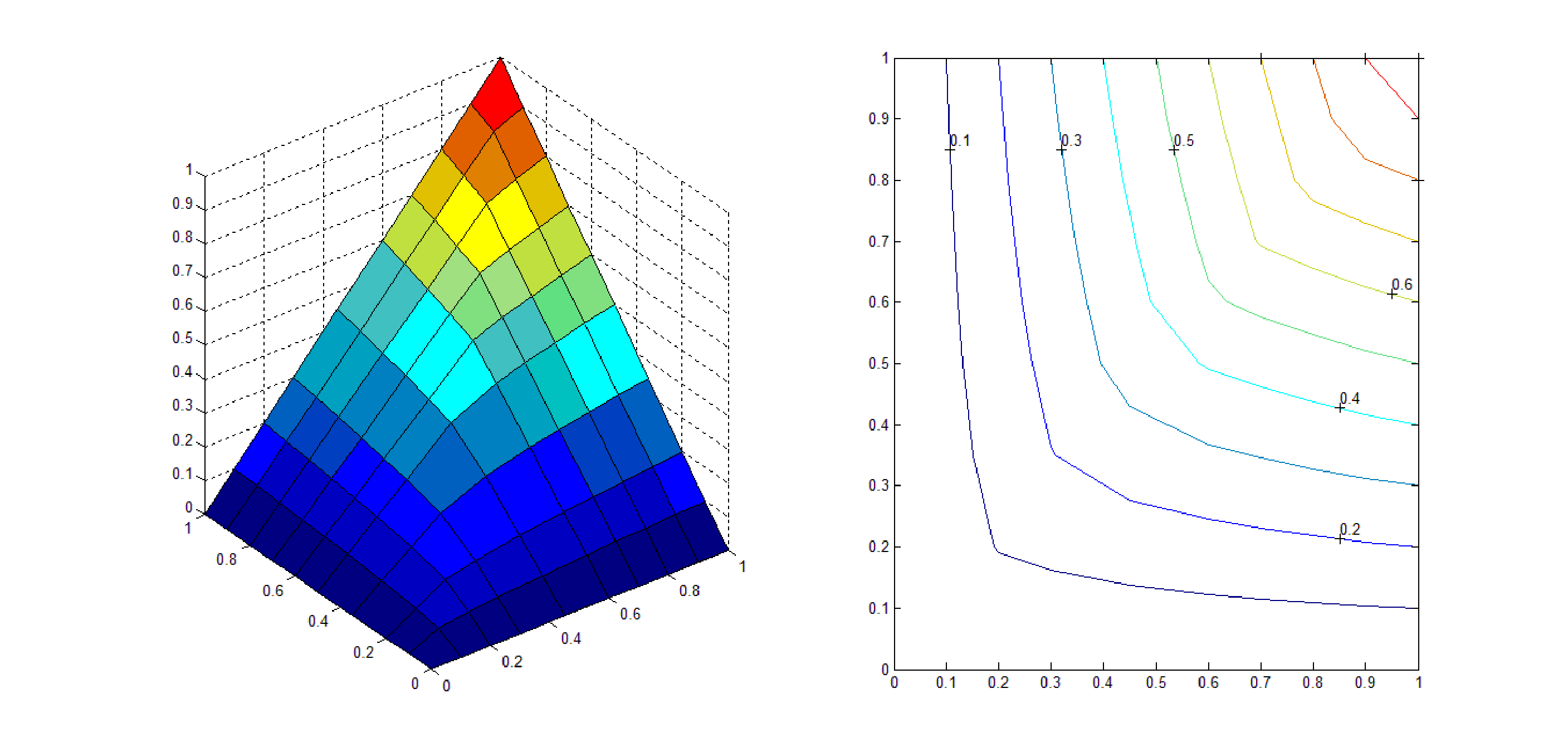}
\end{figure}\end{center}
In order to identify the copula that best fits the data, we need to choose a criterion. As the models are non-nested, following \cite{Rod} we use the Akaike Information Criterion adjusted for small sample bias (\cite{Br})
\[AIC=2k-2l(\widehat{\theta})+\frac{2k(k+1)}{n-k-1}\]

where $l(\widehat{\theta})$ is the maximised log likelihood function, $k$ is the number of estimated
parameters, and $n$ is the sample size. According to this criterion,
the best fitting model is the one that minimises \emph{AIC}. A goodness of fit (\emph{GOF}) test based on the empirical copula \cite{De} is also performed in order to select the best fit copula. Generally, the smallest distance between the empirical copula and the reference copula implies the best fit. Such a distance is measured by the Cram\'{e}r von Mises statistic (\cite{Fe} and \cite{Ge}). Following \cite{Ge}, we use a parametric bootstrap procedure to
obtain approximate p-values\footnote{All the methods discussed
in this article are implemented in the Matlab-program. They are available on request to the authors.}

\begin{table}\caption{Goodness of fit mesures}\label{tab2}
\begin{center}\begin{tabular}{|c|c|c|c|c|c|c|}
\hline              & \multicolumn{2}{|c|}{\emph{IT-UK}}            &\multicolumn{2}{|c|}{\emph{IT-DE}}&      \multicolumn{2}{|c|}{\emph{UK-DE}}\\\hline
\emph{Copula} & AIC&\emph{GOF test p-value}& AIC&\emph{GOF test p-value}& AIC&\emph{GOF test p-value}\\
\hline \hline
Gaussian      & -4.32    &                      0.54&   -10.3   &          0.50        & -9.18   &    0.52                   \\\hline
Gumbel        & -19.20   &                      0.73&   -22.33 &           0.74         & -18.45  &    0.71             \\\hline
F+C+G & -24.87   &                      0.78&   -25.98 &           0.82         & -18.45  &    0.79                 \\\hline
MO copula           & -34.44   &                      0.91&   -37.89 &           0.94         &  -21.67 &    0.82             \\

\hline
\end{tabular}\end{center}
\end{table}

From Table \ref{tab2}, the copula that best fits the data according to both the AIC and and the GOF test is the MO copula. Furthermore, in order to use the data of non-distressed banks, we apply a censored sampling as showed in Figure \ref{figurea}. The results in Table \ref{tab3} show that the estimates of the parameter $\theta$ increase if we apply a censored sampling. This means that the idiosyncratic becomes more important for all the pairs of countries when we consider the characteristics of all the sample. As $\theta$ is the upper tail dependence parameter, the most important result of this empirical evidence is that the intensity of the upper tail dependence increases by applying a censored sampling. In other words, the contagion risk could be underestimated if we do not consider the characteristics of non-distressed banks.

\begin{table}\label{tab3}\caption{Copula parameters estimates: complete and censored sampling}
\begin{center}\begin{tabular}{|c|c|c|c|}
\hline              & \multicolumn{1}{|c|}{\emph{Italy vs Uk}}            &\multicolumn{1}{|c|}{\emph{Italy vs German}}&      \multicolumn{1}{|c|}{\emph{Uk vs German}}\\\hline
sample & parameter estimate& parameter estimate& parameter estimate\\
\hline
complete sample & $\theta=0.45$& $\theta=0.55$& $\theta=0.31$\\
censoring sample & $\theta=0.76$& $\theta=0.83$& $\theta=0.50$\\
\hline
\end{tabular}\end{center}
\end{table}

\section{Conclusion}

In this paper we propose a novel copula-based approach for modeling systemic risk. In particular, the Marshall-Olkin copula is used to estimate the dependence between times to bank failures located in two different countries.
The main advantage of this model is that the impact of the idiosyncratic and systematic components on the systemic risk can be measured. We highlight that the idiosyncratic component is represented by the continuous part of the copula, the systematic by the discrete part. In order to include in the estimates the information from non-distressed banks, a method for censored sampling is proposed. We propose the maximum likelihood method to estimate the MO copula parameter for both the complete and censored samples.
The proposals are applied to data from European banking systems. The first important result of this empirical analysis is that the MO copula  is the copula that best fits the data according to different goodness-of-fit measures. Furthermore, applying censored techniques increases the impact of the systematic component on the systemic risk. For this reason, we hope to provide a method that central banks can use to supply accurate estimate of contagion risk.

\section{APPENDIX}
\label{App1}

We suggest the maximum likelihood estimator (\ref{estimator}) in the case of type I censored sample. To obtain it, we considered the censored sample as described in Figure \ref{figurea}.
We apply the logit transformation $\theta =(1+\exp (-\psi
))^{-1}$ to the conditional log-likelihood function (\ref{eq4}), so we obtain
\begin{eqnarray*}
l(\psi \vert \hat {u},\hat {v}) &=&k+(m_1 +m_2 +r+s)\ln
[1-(1+\exp(-\psi ))^{-1}]+m_3 \ln [(1+\exp(-\psi
))^{-1}]+\\&-&  (1-(1+\exp(-\psi ))^{-1}) (S_1+S_2)- (1+\exp(-\psi
))^{-1}S_{\max}
\end{eqnarray*}
where
\[S_{1} =\sum_{i=1}^{m+r} [-\ln (\hat {u}_i
)]+(n-m-r)t^*,\]
\[S_{2} =\sum_{i=1}^{m+s} [-\ln (\hat {v}_i)
]+(n-m-s)t^*\] and
$S_{\max } =\sum_{i=1}^m \max [-\ln (\hat {u}_i
),-\ln (\hat {v}_i )]+r t^*+s t^*+(n-m-r-s)t^*$.\\
The previous equation can be simplified and it becomes
\begin{eqnarray*}
l(\psi \vert \hat {u},\hat {v}) &=&k+(m_1 +m_2 +r+s)(-\psi)-(m+r+s) \ln [(1+\exp(-\psi
))]+\\&-& \frac{\exp(-\psi)} {(1+\exp(-\psi))} (S_1+S_2)- (1+\exp(-\psi
))^{-1}S_{\max}
\end{eqnarray*}
By differentiating the log-likelihood function with
respect to $\psi$, we obtain
\begin{eqnarray*}\frac{\partial l(\psi \vert \underline{\hat {u}},\underline{\hat {v}})}{\partial \psi} &=& -(m_1+m_2+r+s)+(m+r+s)\frac{\exp(-\psi)}{(1+\exp (-\psi ))}+\\&-& \frac{\exp(-\psi)}{(1+\exp (-\psi ))^2}S_{max}+(S_1+S_2)
\left[\frac{\exp(-\psi)}{(1+\exp(-\psi))^2}\right] \end{eqnarray*}
Setting $\frac{\partial l(\psi \vert \underline{\hat {u}},\underline{\hat {v}})}{\partial \psi}=0$ we obtain
\[m_3\exp(-2\psi)-(m+r+s-2m_3+S_{\min})\exp(-\psi)-(m+r+s-m_3)=0,\]
where $S_{\min}=S_1+S_2-S_{\max}$.

By solving the previous equation with respect to $\exp(-\psi)$, we obtain two solutions
\[z_{1,2}=\frac{m+r+s-2m_3-S_{\min}\pm \sqrt{(m+r+s-2m_3-S_{min})^2+4m_3(m+r+s-m_3)}}{2m_3}\]
Since only the solution $z_{1}=\frac{m+r+s-2m_3-S_{\min}+ \sqrt{(m+r+s-2m_3-S_{min})^2+4m_3(m+r+s-m_3)}}{2m_3}$  has positive values, it is the unique accepted solution for $\exp(-\psi)$. Hence, the unique solution of the optimization problem is
\[
\hat {\psi }_c =-\ln \left[ {\frac{m+r+s-2m_3 -S_{\min } +\sqrt
{(m+r+S_{\min } -2m_3 )^2+4m_3 (m+r+s-m_3 )} }{2m_3 }} \right].
\]
By checking the sign of the second derivative, we obtain that the previous solution is a maximum.

\end{document}